\pdfoutput=1
\documentclass[onecolumn,pdflatex]{SSA-TSR}

\newcommand{\EtaThresh}{{-1.822}}

\citethisauthor{Sedghizadeh, M., R. Shcherbakov, and M. van den Berghe}
\vol{0}
\iss{0}
\doi{00.0000/000000000}
\recdate{00 Month 0000}

\begin{document}

\title{Generalized nearest-neighbor distance and Hawkes point process modeling applied to mining induced seismicity}

\author[1\orc{0000-0003-3713-3782}]{Mohammadamin Sedghizadeh}
\author[*1,2\orc{0000-0002-3057-0851}]{Robert Shcherbakov}
\author[3]{Matthew van den Berghe}

\affil[1]{Department of Earth Sciences, University of Western Ontario, London, ON, Canada, N6A 5B7.}{\auorc[https://orcid.org/]{0000-0003-3713-3782}{(MS)}}
\affil[2]{Department of Physics and Astronomy, University of Western Ontario, London, ON, Canada, N6A 3K7.}{\auorc[https://orcid.org/]{0000-0002-3057-0851}{(RS)}}
\affil[3]{Department of GeoServices, Nutrien Ltd., Saskatoon, SK, Canada, S7K 5R6.}
{}
\corau{*Corresponding author: rshcherb@uwo.ca}

\begin{abstract}
    Modeling seismic activity rates and clustering plays an important role in studies of induced seismicity associated with mining and other resource extraction operations. This is critical for understanding the physical and statistical characteristics of seismicity and assessing the associated hazard. In this work, we introduce the generalization of the Nearest-Neighbor Distance (NND) method by incorporating an arbitrary distribution function for the frequency-magnitude statistics of seismic events. Operating within a rescaled hyperspace that includes spatial, temporal, and magnitude domains, the NND method provides an effective framework for examining seismic clustering. By integrating a mixture of the two tapered Pareto distributions, the generalized NND approach accommodates deviations from standard frequency-magnitude scaling when studying the clustering properties of seismicity. In addition, the application of the temporal Hawkes process to model the mining seismicity rate reveals that the seismicity is primarily driven by external factors and lacks pronounced interevent triggering. A case study from a potash mine in Saskatchewan is presented to illustrate the application of the generalized NND method and the Hawkes process to estimate the clustering properties and occurrence rates of induced microseismicity. The implications of observed temporal variations and clustering behavior are discussed, providing insights into the nature of induced seismicity within mining environments.
\end{abstract}

\maketitle

\section{Introduction}

Induced seismicity, the occurrence of seismic events that typically accompany energy/mineral resource extraction operations such as mining, hydraulic fracturing and waste water injection, reservoir impoundment, and others, has become a critical area of research due to its implications for operational safety, infrastructure integrity, and seismic hazard assessment \citep{Ellsworth13a,GrigoliCPR17a,AtkinsonEI2020a,SchultzSBEBE20a}. These seismic events, while typically of lower average magnitudes than natural earthquakes, often exhibit a variety of spatial, temporal, and magnitude patterns that are influenced by anthropogenic factors rather than tectonic plate interactions \citep{AtkinsonEGW16a,KeranenW18a}. For example, mining-induced seismicity is frequently caused by the redistribution of stresses within rock masses as a result of excavation and ore extraction \citep{GendzwillS92a,GibowiczL00a,LasockiO08a,Gibowicz09a}. Understanding the underlying mechanisms and characteristics of induced seismicity is essential for mitigating associated risks, improving forecasting models, and developing effective hazard assessment frameworks \citep{vanderElstPWGH2016a,ShcherbakovTRTH10a,ShcherbakovTR15a,LangenbruchWZ2018a,SchultzSBEBE20a,ZhouLGSC2024a}. To accomplish these tasks the appropriate statistical models need to be formulated and used. 

In statistical seismology, the analysis of natural seismicity is typically based on the application of several well established statistical models and approaches. This includes the application of the Gutenberg-Richter (GR) law to model the frequency-magnitude statistics of earthquakes \citep{GutenbergR44a}. In many instances, the analysis of the seismicity rates is performed using the temporal or spatio-temporal Epidemic Type Aftershock Sequence (ETAS) model \citep{Ogata88a,Ogata98a,Ogata2024a}. The clustering aspects of seismicity can be studied by using the Nearest-Neighbor Distance (NND) analysis \citep{BaiesiP04a,ZaliapinGKW08a,ZaliapinB13a,ZaliapinB2022a}. These models are typically used in a Probabilistic Seismic Hazard Analysis (PSHA) to forecast future seismic activity \citep{LlenosM13a,LlenosM20a,MizrahiNCW2024a,LlenosMSR2024a}. 

While effective for natural seismicity, the above mentioned statistical models are not always applicable to some instances of induced seismicity. Deviations from GR scaling are commonly observed, necessitating the formulation of alternative statistical models \citep{EatonDPB2014a,UrbanLBNGK2016a,IgoninZE18a,Leptokaropoulos2020a,HerrmannPM2022a,SedghizadehBS2023a,KrushelnitskiiSVSA2024a}. While the ETAS model has proven effective for analyzing natural aftershock sequences, its applicability to induced seismicity is limited due to the absence or minimal presence of pronounced aftershock patterns commonly observed in tectonic environments. However, in some instances of mining seismicity, it was shown that events can generate aftershock sequences that follow the modified Omori law \citep{KgarumeSD10a,VallejosM10a,EstayVPB2020a}. Meanwhile, the ETAS model has been applied to some cases of induced mining seismicity to forecast its evolution \citep{MartinssonT20a,GospodinovDD2022a,SedghizadehBS2024a}. While both the modified Omori law and ETAS models are effective for sequences driven by magnitude scaling, they are less suited for cases where external factors, such as operational activities, primarily govern seismicity rates.

Beyond frequency-magnitude statistics, the study of seismic clustering -- where events occur closely in space and time -- is equally critical for understanding seismic dynamics \citep{SchoenballDG15a,CochranRHDR2018a,LiSR2021a}. Clustering provides insights into the underlying stress redistribution processes and helps differentiate between background seismicity and triggered events. For example, induced seismicity often exhibits unique spatiotemporal clustering patterns, reflecting the influence of human activities rather than natural tectonic processes \citep{SchoenballE17a,CochranWKH2020a,KarimiD2023a,Li2025a}. Among the tools adapted to study clustering aspects of natural and induced seismicity, the NND method can play a critical role \citep{ZaliapinB2022a}. However, the standard NND method relies on the GR law to describe the frequency-magnitude distribution of earthquakes, which often fails to capture the statistical properties of mining-induced seismicity.

The temporal Hawkes point process, of which the ETAS model is a special case, provides a mechanism for capturing the stochastic nature of the occurrence of random events. Introduced by \citet{Hawkes1971a}, this model quantifies the likelihood of an event triggering subsequent events, with its conditional intensity function reflecting the influence of past events. This approach can be well-suited for examining induced seismicity and has applications in fields ranging from financial transactions and social network dynamics to solar physics \citep{ZipkinSCB2016a,Ross2020a,Schoenberg2023a,BernabeuZM2024a}. In a recent study, a nonparametric Hawkes model has been developed to incorporate spatial inhomogeneity, anisotropy, and space-time non-separability, better reflecting the heterogeneous geological conditions observed in seismic data \citep{KwonZJ2023a}. In the Hawkes point process, the event magnitudes do not influence the occurrence of subsequent events and this mechanism can explain some cases of induced seismicity.

In this work, we perform the analysis of mining induced microseismicity in a soft rock environment by integrating several statistical methods. First, we reformulate the NND method by incorporating an arbitrary distribution function for the frequency-magnitude statistics of event magnitudes, enabling a more accurate and nuanced analysis of seismic clusters. This generalization overcomes a limitation of the standard NND method, particularly its dependence on the GR law, which often fails to adequately capture the statistical properties of induced seismicity. By accounting for deviations from natural seismicity patterns, the generalized NND method provides a robust framework for identifying and characterizing clustering behavior specific to mining-induced seismic events. We also introduce a mixture model to describe the statistics of earthquake magnitudes that takes into account a bimodal nature of some instances of mining or induced seismicity. Furthermore, we use the temporal Hawkes point process to quantify the observed event rate and investigate interactions between background and triggered seismicity. By integrating these approaches, our study provides a unified view of the clustering behavior and evolution of mining-induced seismicity. The developed methodology is applied to one particular instance of mining microseismicity in a potash mine in Saskatchewan, Canada.

\section{Mining seismicity catalog} \label{sec:data}

This study focuses on microseismicity in a potash mine located in Saskatchewan, Canada. Mining operations are conducted within the Prairie Evaporate formation, which is the source of potash deposits \citep{Broughton2019a}. This formation has a thickness of 100-200 m and is located at a varying depth of 900-1100 m. The microseismicity is typically detected above the mining horizon in the Devonian carbonate formation, which has approximate thickness of 400-500 m \citep{FunkDML2022a,BarthwalBS2024a}. The main factors that drive the occurrence of microevents is the changes in the local stress field due to excavation of ores and time dependent deformation of the overlying strata without presence of any preexisting fault structures \citep{SepehrS88a,HasegawaWG89a,GendzwillS92a,FunkILB19a,BarthwalS2024a}.

\begin{figure}[!ht]
\centering
{\includegraphics*[width=0.8\linewidth, trim=15mm 50mm 7mm 53mm]{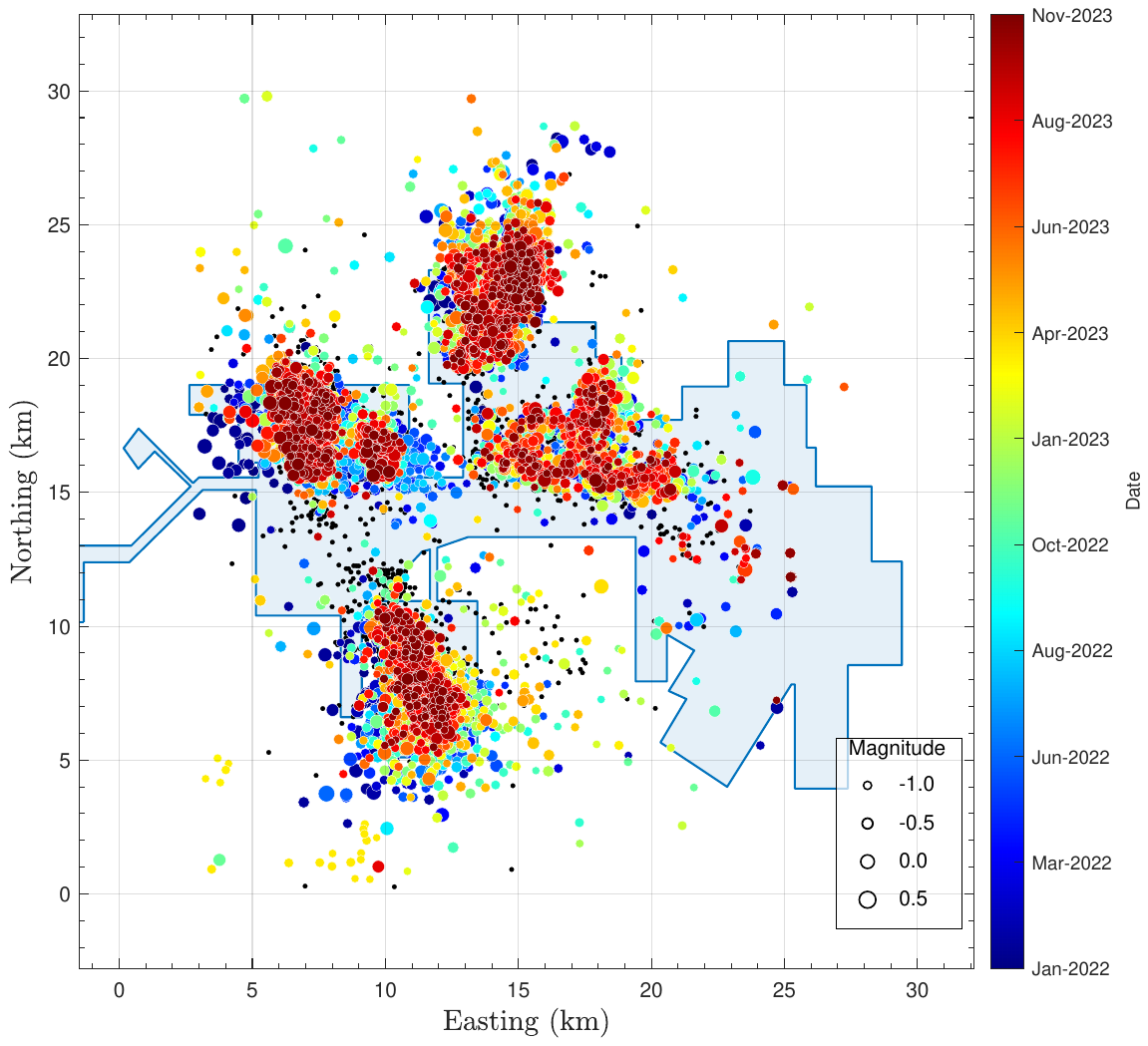}}
\caption{Microseismic event epicenters from a potash mine in Saskatchewan. The catalog spans the time interval from January 1$^\mathrm{st}$, 2022, to November 15$^\mathrm{th}$, 2023. The colored solid circles show events with $m \geq -1.0$, while black solid circles represent events with magnitudes below $m < -1.0$. The colors of the circles reflect the  occurrence times of microevents, starting from January 1$^\mathrm{st}$, 2022, as given by the color bar. The light blue area illustrates the mine layout.}
\label{fig1}
\end{figure}

The microseismicity catalog analyzed covers the period from January 1$^\mathrm{st}$, 2022, to November 15$^\mathrm{th}$, 2023. To preprocess the catalog, spatial boundaries were applied to exclude events occurring outside the mining area, and an unsupervised clustering approach, specifically the $K$-means algorithm, was used to reduce false-positive detections. In particular, hyperspace clustering was utilized to identify and remove falsely detected events in the southern and southeastern parts of the mine, which were triggered by dumping ore rocks into a large, deep pit. Following these steps, the final catalog for the study time interval consists of 20,977 events with magnitudes ranging from $-3.28$ to $0.55$. In Figure~\ref{fig1}, the epicentres of the mining events are plotted on top of the mine layout. The microseismic events are reported up to 2.1 km deep below surface.

\section{Statistical methods}
\label{sec:methods}

\subsection{A generalized Nearest-Neighbor Distance method}
\label{methods_ndd}

Seismic events are typically triggered by natural or anthropogenic changes in the stress field, often leading to clustering in seismic sequences. The NND method is a powerful tool for studying such clustering by analysing rescaled relative distances between all pairs of earthquakes within a space formed by temporal, spatial, and magnitude dimensions. This method, introduced by \citet{BaiesiP04a,ZaliapinGKW08a}, minimizes the rescaled distances between seismic events, providing information on the clustering behavior of seismicity. One limitation of the standard NND method is its reliance on the assumption that seismic events follow the GR distribution.

Here, we consider a generalization of the standard NND approach. In a seismic catalog, each event $i$ is reported by providing its time of the occurrence $t_i$, magnitude $m_i$, and epicenter location (latitude $\phi_i$, longitude $\lambda_i$). For any pair of events $i$ and $j$, the rescaled distance $\eta_{ij}$ in spatio-temporal and magnitude domains is computed as:
\begin{equation}\label{eq1}
    \eta_{ij}  = \left\{
    \begin{array}{lc}
      t_{ij}\, r_{ij}^{d_f}\, f(m_i)\,, & \quad \textrm{for} \quad t_{ij} > 0\,, \\
      \infty\,,                               & \quad \textrm{for} \quad t_{ij} \leq 0\,, 
    \end{array}
    \right.
\end{equation}
where $ t_{ij} = t_j - t_i $ is the time interval between pairs of events $i$ and $j$ and $r_{ij}$ is the epicentral distance. In the generalization of the NND method, we assume that the weight of the parent event magnitude, $f(m_i)$, is given by a probability density function for event magnitudes. In the standard NND method the weighting factor is $f(m_i) = 10^{-b m_i}$ \citep{ZaliapinGKW08a}, where the $b$-value is estimated from the GR relation. The parameter $d_f$ represents the fractal dimension of the epicenter distribution, which is found between 1.2 and 1.8 for earthquake distributions in two dimensions \citep{KosobokovM94a,ZaliapinB13a}. 

The rescaled distance $\eta_{ij} = T_{ij} \cdot R_{ij}$ represents the combined influence of temporal and spatial separation between pairs of seismic events, weighted by their magnitudes, where $T_{ij}$ is the rescaled temporal component, and $R_{ij}$ is the rescaled spatial component. These are defined as follows:
\begin{eqnarray}
    \label{eq2}
    T_{ij} & = & t_{ij} \, \left[f(m_i)\right]^{q}\,, \\
    \label{eq3}
    R_{ij} & = & r_{ij}^{d_f} \, \left[f(m_i)\right]^{p}\,,
\end{eqnarray}
in which the exponents $q$ and $p$ satisfy $p+q=1$ and determine the relative influence of the $i$th event magnitude on the rescaled time and distance, respectively. Typically, $q = p = \frac{1}{2}$, ensuring that the magnitude influence is evenly distributed between time and space. For each event $j$, the minimum value $\eta_j = \min\limits_{i:i < j} (\eta_{ij})$ represents the \emph{nearest-neighbor proximity distance}, with event $i$ considered the parent of event $j$. By analyzing the distribution of nearest-neighbor proximities $\eta$, as well as the rescaled time ($T$) and spatial distance ($R$) distributions, one can gain insight into the spatiotemporal clustering of seismic events.

To apply this model to instances of induced seismicity, we work in the moment domain and consider several standard distributions and their mixtures. The moment is related to the magnitude scale as $M(m) = 10^{\frac{3}{2}(m+10.73)}$. The equivalent of the exponential distribution (GR scaling) in the moment domain is the Pareto distribution \citep{Kagan02a}. The probability density function (pdf) of the Pareto distribution is:
\begin{equation} \label{eq4}
    f_\mathrm{Par}(M) = \beta \, M_\mathrm{min}^{\beta} M^{-1-\beta}\,,
\end{equation}
where $\beta$ is the exponent related to the $b$-value: $b = \frac{3}{2}\,\beta$ of the GR law and $M_\mathrm{min}$ is the lower moment cutoff. It was also suggested by \citet{Kagan02a} to consider the tapered Pareto distribution which has the following pdf:
\begin{equation} \label{eq5} 
    f_\mathrm{tap}(M)= \left( \frac{M_\mathrm{min}}{M} \right)^{\alpha} \left[ \frac{{\alpha}}{M} + \frac{1}{M_\mathrm{cm}} \right] \exp \left( \frac{M_\mathrm{min} - M}{M_\mathrm{cm}} \right)\,,
\end{equation}
where $\alpha$ and $\beta$ are the shape parameters for the tapered Pareto distribution and $M_\mathrm{cm}$ specifies the corner moment.

Induced seismicity may exhibit a bimodality in the frequency-magnitude distribution reflecting the presence of distinct populations of seismic events. This can be effectively modeled by using a mixture of distributions:
\begin{equation} \label{eq6}
    f_\mathrm{mix}(M) = w \cdot f_\mathrm{1}(M) + (1 - w) \cdot f_\mathrm{2}(M)\,,
\end{equation}
where $w$ is a mixing weight, which determines the relative contribution of the two distributions $f_1(M)$ and $f_2(M)$. For example, one can use the mixture of the Pareto or tapered Pareto distributions, each characterizing a specific subset of seismic events. The parameters of the mixture model, including the mixing weight $w$ and the parameters of $f_1(M)$ and $f_2(M)$, can be estimated using the Maximum Likelihood Estimation (MLE) method.

The mixture model (\ref{eq6}) can be incorporated into the NND analysis, equation~(\ref{eq1}), by converting from the moment to the magnitude scale: $f(m) = \frac{3}{2}\log(10)\,f_\mathrm{mix}(M(m))\,10^{\frac{3}{2}(m+10.73)}$, where the relation between the moment and magnitude scales is used. This modification allows the NND method to capture non-standard frequency-magnitude characteristics observed in some instances of induced seismicity.

\subsection{The temporal Hawkes point process} \label{methods_hawkes}

The conditional intensity function of the Hawkes point process, $\lambda_\omega (t|\mathcal{H}_t)$, defines the instantaneous rate of occurrence of events at time $t$ given the history of past seismicity, $\mathcal{H}_t$, which includes the times of all previous events above a given lower magnitude cutoff $m_0$. This is given as \citep{Hawkes1971a}:
\begin{equation} \label{eq7}
    \lambda_\omega (t|\mathcal{H}_t) = \mu + A \sum_{i:t_i<t}^{N_t} e^{-\alpha (t-t_i)}
\end{equation}
where $\mu$ represents the background rate, which is the baseline occurrence rate of events that is independent of interevent triggering. The parameter $A$ quantifies the productivity of prior events. Larger values of $A$ indicate a stronger potential for an event to generate subsequent events. The parameter $\alpha$ controls the rate at which the influence of past events decays over time; larger values of $\alpha$ lead to faster decay, meaning that recent events exert a stronger influence on the current occurrence rate than older ones. Here, $t_i$ denotes the occurrence times of past events, where $t_i < t$, and $N_t$ is the total number of events up to time $t$.

\section{Results} \label{sec:results}

\subsection{The magnitude of completeness $m_\mathrm{c}$ and the frequency-magnitude statistics}
\label{mag_compl}

The magnitude of completeness, $m_\mathrm{c}$, can be determined by using the Goodness-of-Fit (GoF) criterion \citep{WiemerW00a}. In this method, $m_\mathrm{c}$ is identified by comparing the observed frequency-magnitude distribution with a theoretical model by varying the cutoff magnitude. The GoF value quantifies the discrepancy, with $m_\mathrm{c}$ defined as the cutoff magnitude where the GoF exceeds a specified threshold. Model parameters are simultaneously estimated, and residuals are calculated based on the absolute differences between the observed and modeled numbers of events in each magnitude bin \citep{WiemerW00a}. 

\begin{figure}[!ht]
\centering
{\includegraphics*[width=0.85\linewidth, trim=10mm 65mm 15mm 70mm]{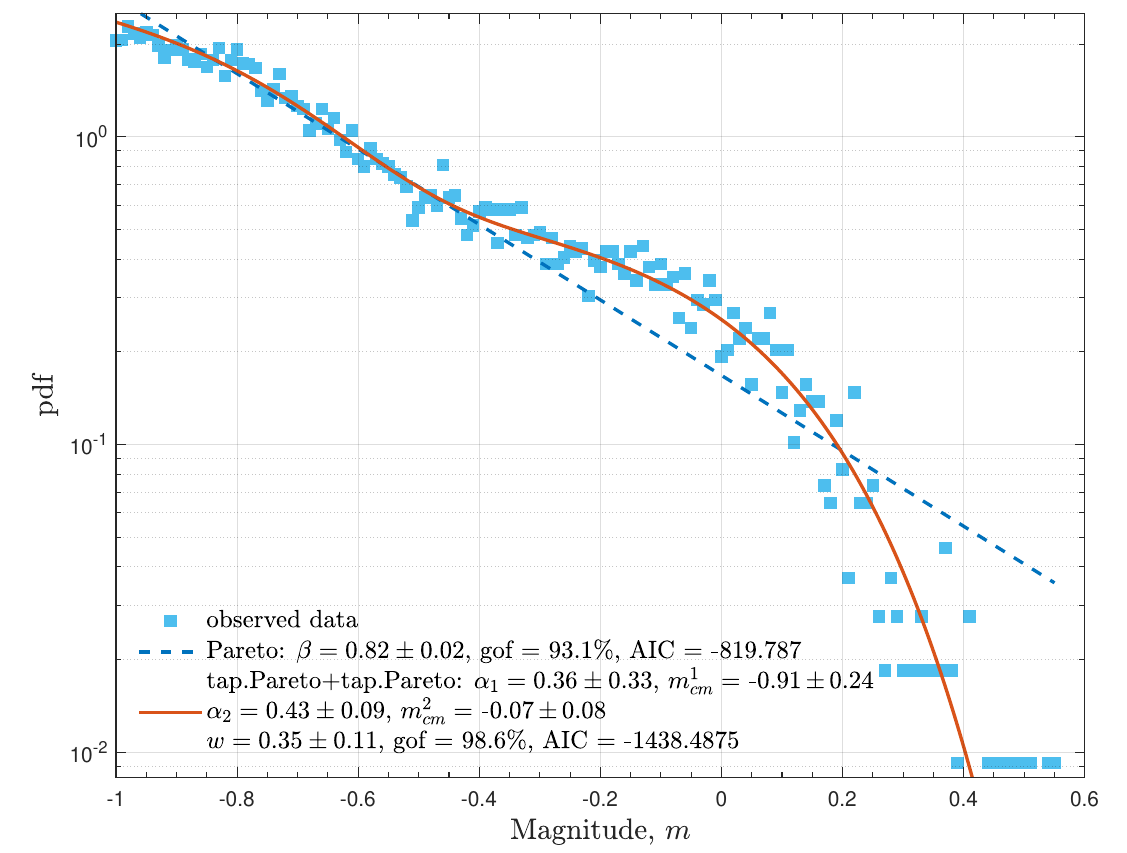}}
\caption{Modeling the frequency-magnitude statistics of the mining microseismicity. The solid blue squares indicate the normalized numbers of the observed event magnitudes in each magnitude bin. The fit of the Pareto distribution~(\ref{eq4}) is shown as a dark-blue dashed line. The fit of the mixture~(\ref{eq6}) of the two tapered Pareto distributions~(\ref{eq5}) is plotted as an orange curve. The estimated parameters for the both models are reported in the legend.}
\label{fig2}
\end{figure}

The observed frequency-magnitude statistics of microevents displays a bi-modal structure (Figure~\ref{fig2}). The statistics of magnitudes was modelled by using four distributions: the Pareto distribution (\ref{eq4}); the tapered Pareto distribution (\ref{eq5}); the mixture, as defined in (\ref{eq6}), of the Pareto (\ref{eq4}) and tapered Pareto (\ref{eq5}) distributions; and the mixture of the two tapered Pareto distributions. These models were defined in the seismic moment domain by transforming the magnitudes of the events into the moments. Based on the results of the GoF (Figure~S1), we estimated the lower magnitude cutoff $m_\mathrm{c} = -1.0$, above which the catalog is assumed to be complete. The total number of events above magnitude $m\geq -1.0$ and during the time interval $[T_s\,,T_e]=[59,\,683]$ days was 10,858. Using this lower magnitude cutoff $m_\mathrm{min} = -1.0$, we evaluated the fits of those four distributions. The estimated model parameters and the corresponding 95\% confidence intervals, are given in Table~S1. Among the tested models, the frequency-magnitude statistics of microseismicity were best described by the mixture of the two tapered Pareto distributions, shown in Figure~\ref{fig2}, as indicated by the Akaike information criterion. The equivalent fits in the seismic moment domain are presented in Figure~S2. In addition, the fits of the all four models are given in Figures~S3 and S4.

\subsection{Application of the generalized NND method}

\begin{figure}[!ht]
\centering
{\includegraphics*[width=0.8\linewidth, trim=38mm 14mm 40mm 10mm]{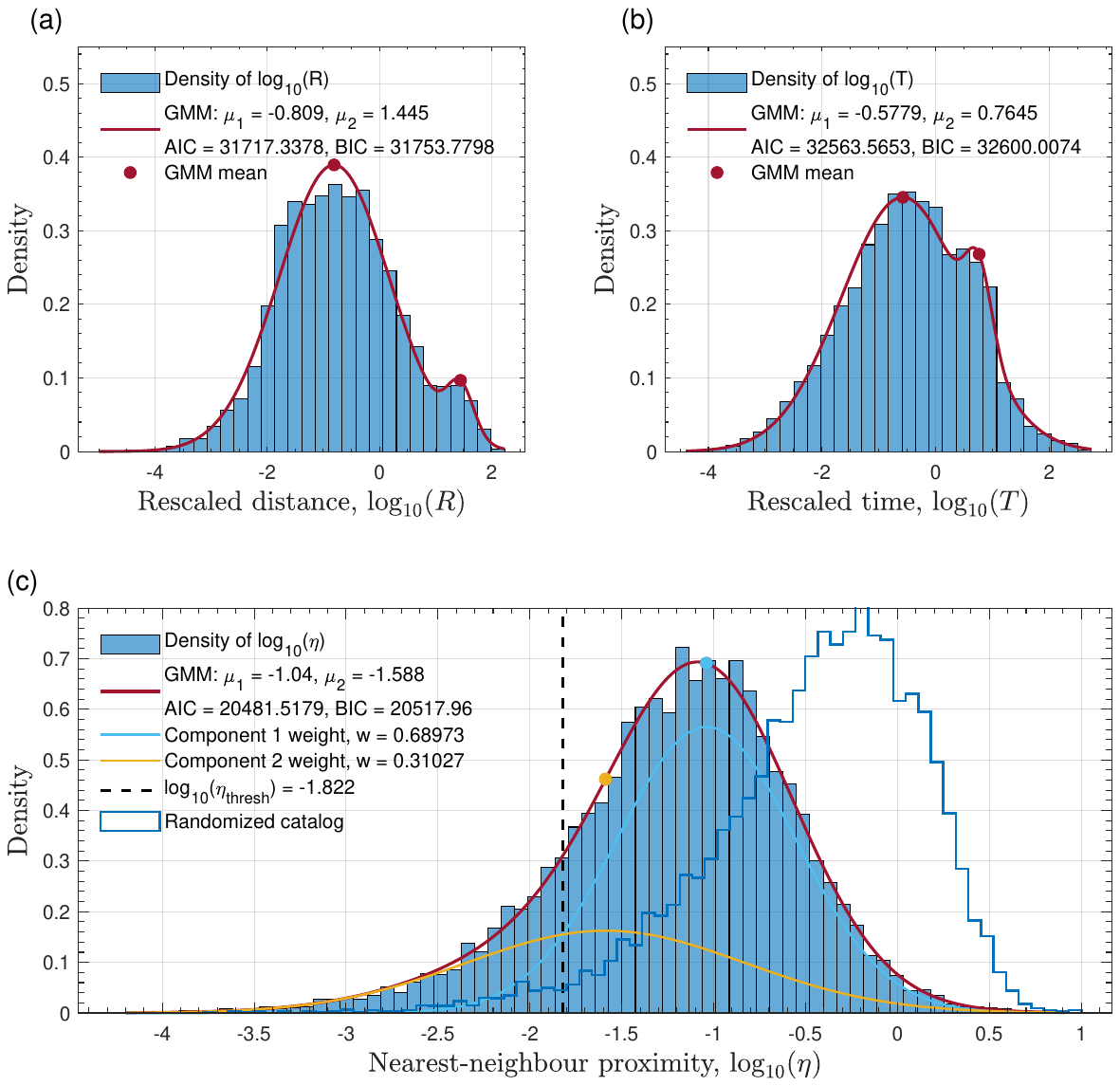}}
\caption{Empirical distributions of (a) the rescaled distances $R$; (b) the rescaled times $T$; and (c) the nearest-neighbor proximities $\eta_{j}$. All microevents above magnitude ${m \geq -1}$ were used. A two-component GMM is used to fit each distribution in (a)-(c) and the GMM fits are plotted as dark red curves. (c) The individual components of the GMM in case of the distribution of the nearest-neighbor proximities are plotted as yellow and blue solid curves. The parameters of the GMM are reported in the legend. The intersection of the two components of the GMM is used to compute the threshold ${\log_{10}(\eta_\mathrm{thresh}) = \EtaThresh}$ and is shown as a vertical black dashed line. The distribution of $\eta$ for a fully randomized in space and time catalog is illustrated as a blue stairstep plot.}
\label{fig3}
\end{figure}

The standard NND method assumes that the frequency-magnitude distribution of events follows the GR scaling \citep{ZaliapinB13a}. To add flexibility in seismic cluster analysis, we generalized the NND method by considering an arbitrary distribution of event magnitudes $f(m)$ in equations (\ref{eq1})-(\ref{eq3}). From the analysis of the frequency-magnitude statistics of the mining microseismicity, we obtained that the mixture model (\ref{eq6}) consisting of the two tapered Pareto distributions (\ref{eq5}) fits the data best. Therefore, we incorporated this mixture model into the NND analysis. Figure~\ref{fig3} shows the computed distributions for $R$, $T$, and $\eta$ using the generalized NND method (\ref{eq1})-(\ref{eq3}) with the parameters for the mixture of the two tapered Pareto distributions given in Table~S1 and the fractal dimension, $d_f = 1.6$. 

To model the distributions of the rescaled nearest-neighbor proximities $\eta$, the rescaled times $T$, and the rescaled distances $R$, we used the Gaussian Mixture Model (GMM) with two components. The corresponding fits are given in Figure~\ref{fig3}. We considered several GMM components ranging from 1 to 4 and the two-component model achieved the best fit. The rescaled proximities $\eta$ do not show a distinct separation into two modes as it is typically observed for tectonic seismicity \citep{ZaliapinB13a}. However, one still can define a separation between the two modes by considering the intersection point at ${\log_{10}(\eta_\mathrm{thresh}) = \EtaThresh}$ between the two Gaussian modes. This is illustrated as a vertical dashed line in Figure~\ref{fig3}c. We used this threshold value to separate the microseismicity into background events that lie to the right from the threshold and clustered events that lie to the left. To illustrate the effects of the spatial and temporal organization of seismicity on the distribution of $\eta$, we performed the NND analysis on the randomized in space and time catalogue. This is shown as the straistep plot in Figure~\ref{fig3}c. The joint distribution of the rescaled spatial $R$ and temporal $T$ distances is shown in Figure~S5 to illustrate the clustering modes. The separation between the two modes is shown by the dashed white line using the condition $\log_{10}(R) + \log_{10}(T) = \log_{10}(\eta_\mathrm{thresh})$, where the threshold $\log_{10}(\eta_\mathrm{thresh}) = \EtaThresh$ is estimated from the intersection of the two components of the GMM fit.

\begin{figure}[!ht]
\centering
{\includegraphics*[width=0.8\linewidth, trim=40mm 14mm 40mm 16mm]{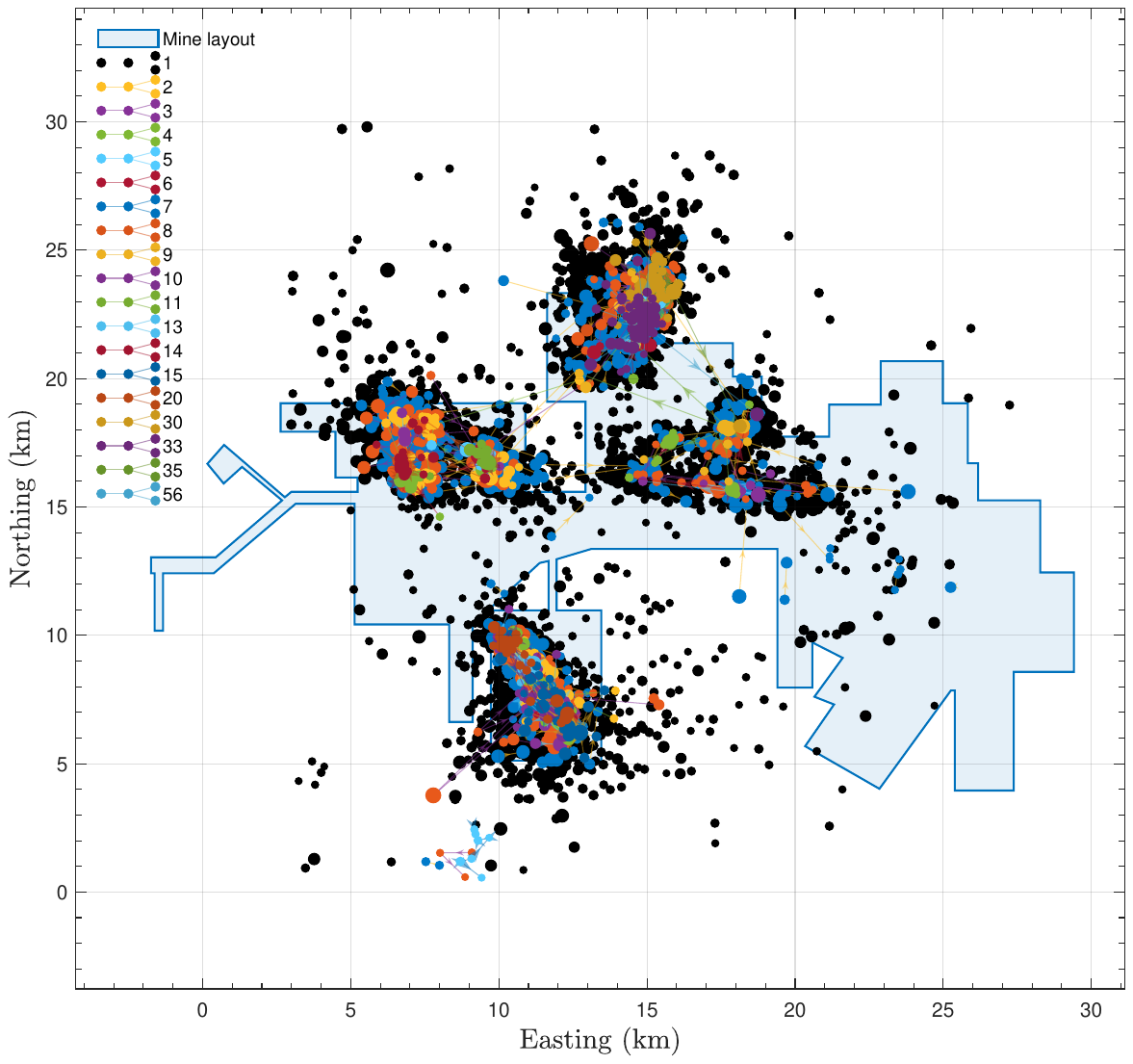}}
\caption{Spatial distribution of the microseismic event clusters that were identified by applying the generalized NND method. The solid circles with varying colors indicate events that belong to clusters with more than one event. The black solid circles show the single events. The threshold ${\log_{10}(\eta_\mathrm{thresh}) = \EtaThresh}$ was used to separate the events.}
\label{fig4}
\end{figure}

After declustering, 9070 events were identified as background events, while the remainder formed clustered events (Figures~S5 and S6). This also resulted in the formation of a total of 1196 family trees with clusters containing more than one event. The NND method subdivided the analyzed sequence into 12 classes (Figure~S6). The first 9 largest clusters are shown in Figure~S7. For each cluster, one can compute the following characteristics: the number of events $N$ in each cluster; the difference in magnitude between the largest (root) and the second largest event $\Delta m$; the cluster duration $T_d$ in days; the average leaf depth $\langle d \rangle$; the normalized leaf depth $\delta = \frac{\langle d \rangle}{\sqrt{N}}$; and the inverted branching number $B_\mathrm{I}$ \citep{ZaliapinB13a,KothariSA20a,SedghizadehBS2023a}. In Figure~\ref{fig4}, we show the spatial distribution of the background events as black solid circles and the clustered events as colored solid circles. 

\subsection{Modeling the microseismicity rate with the temporal Hawkes point process}

\begin{figure}[!ht]
\centering
{\includegraphics*[width=0.8\linewidth, trim=15mm 35mm 5mm 30mm]{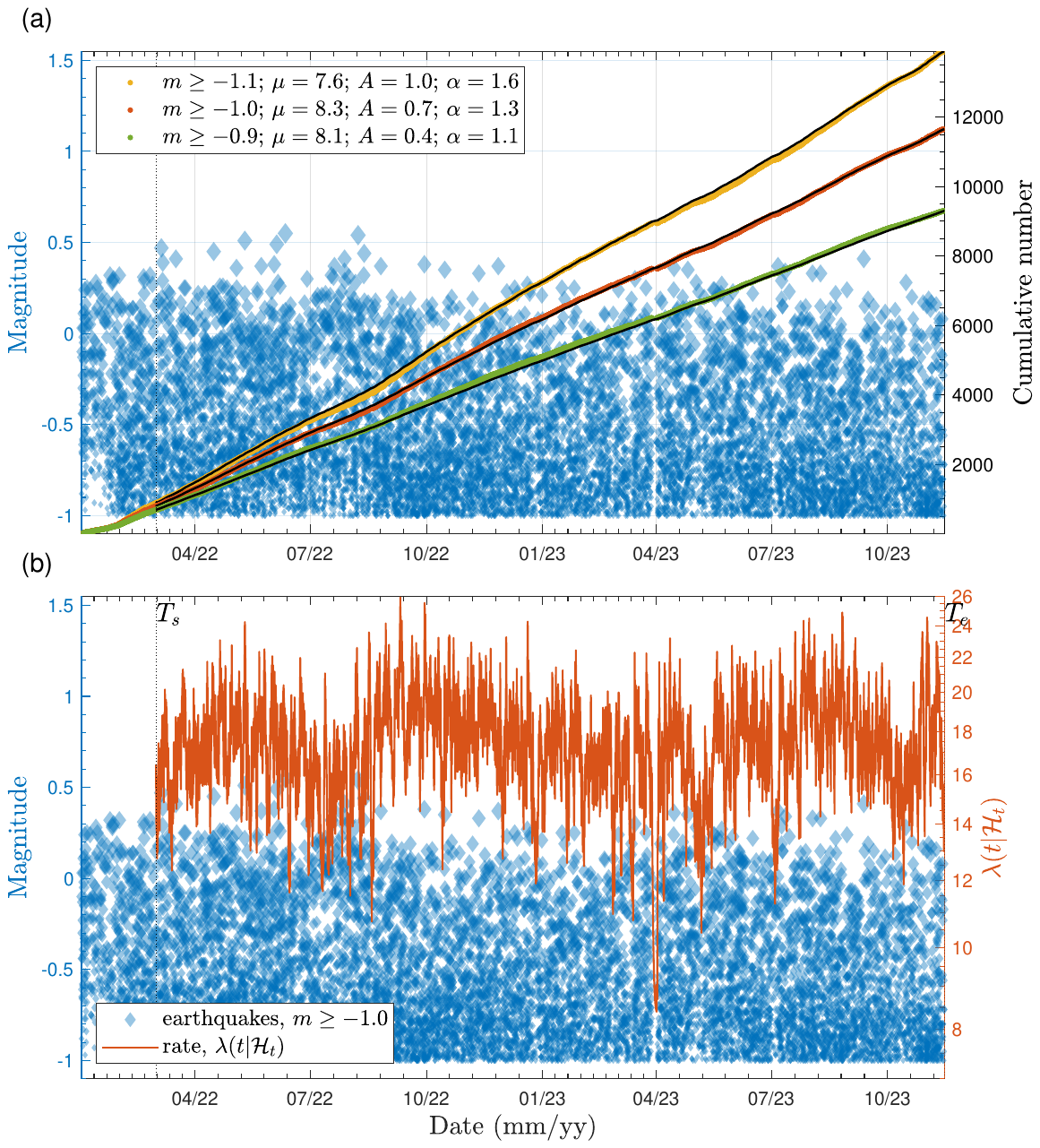}}
\caption{Fit of the Hawkes model, equation~(\ref{eq7}), applied to the mining microseismicity during the target time interval $[T_s,\,T_e ]=[59,\,683]$ days. a) The sequence of event magnitudes during the whole study time interval $[T_0,\,T_e ]=[0,\,683]$ days is shown for events above $m\geq -1.0$. The cumulative number of observed events and the corresponding model fit of the Hawkes process for events above $m\geq -0.9$, $-1.0$, and $-1.1$ are plotted as green, orange, and yellow curves, respectively. The estimated model parameters for each magnitude cutoff are given in the legend. b) The microseismicity sequence and the corresponding fit of the Hawkes model as the conditional intensity function, equation~(\ref{eq7}), are plotted for events above $m\geq -1.0$.}
\label{fig5}
\end{figure}

The temporal Hawkes point process (\ref{eq7}) was applied to model the mining microseismicity rate. The model parameters $\{\mu,\,A,\,\alpha\}$ were estimated by using the MLE method during the target time interval $[T_s\,,T_e]=[59,\,683]$ days. The full length of the catalog in the time interval $[T_0,\,T_e]=[0,\,683]$ days was used to define the rate. We estimated the values of the parameters $\mu=8.3\pm0.5,\,A=0.7\pm0.4,\,\alpha=1.3\pm0.5$ using the lower magnitude cutoff $m_0 = -1.0$. The Hawkes model fits the observed event rate very well, and the results are illustrated in Figure~\ref{fig5}. We also provide the fits of the model for the two additional lower magnitude cutoffs $m_0 = -1.1$ and $m_0 = -0.9$ (Figure~\ref{fig5}a) to illustrate the parameter dependence on $m_0$. It can be seen that with decreasing lower magnitude cutoff $m_0$ the both parameters $A$ and $\alpha$ increase.

\section{Discussion} \label{sec:discussion}

Induced seismicity in a soft rock mining environment can exhibit properties that differ from seismicity observed in tectonic settings. This study models induced microseismicity resulted from potash mining operations by integrating several statistical techniques, specifically the mixture model for the frequency-magnitude statistics of events, the generalized NND method and the Hawkes point process.

Our comparative analysis of several probability distribution functions, including Pareto and tapered Pareto distributions, revealed that a mixture model of the two tapered Pareto distributions approximated the best the observed bi-modal frequency-magnitude statistics of mining microseismicity. Similar observations have been reported in studies of seismicity, where deviations from GR scaling highlight the complexity of anthropogenic seismic processes \citep{UrbanLBNGK2016a,IgoninZE18a}. In addition, the generalized NND method provided a refined framework for analyzing seismic clusters by incorporating the mixture of two tapered Pareto distributions. This method addressed a limitation of traditional NND analysis reliant on GR-based scaling assumptions. By applying a threshold to the rescaled nearest-neighbor proximity distances ($\eta$), we effectively identified clustered events in rescaled temporal and spatial distances (Figure~S5).

The spatial distribution of mining microseismicity revealed a strong concentration near active mining zones (Figure~\ref{fig1}), supporting the hypothesis that mining operations play a direct role in event triggering and clustering. This observation is consistent with the correlation between extraction rates and seismicity rates observed in a different potash mine \citet{SedghizadehBS2023a}, as well as findings from other studies on induced seismicity that highlight similar spatial clustering around operationally active areas \citep{CochranWKH2020a}. These results emphasize the critical importance of location-specific seismic monitoring to identify zones of increased seismic activity and implement strategies to mitigate associated risks.

The application of the temporal Hawkes point process showed steady oscillations in the induced seismicity rate (Figure~\ref{fig5}b). The estimated model parameters revealed a high background rate ($\mu = 8.3$ for events above $m\ge -1.0$), indicating that much of the observed seismicity is governed by background processes rather than self-excited clustering. This result is consistent with other studies of induced seismicity, where operational triggers often dominate over aftershock-like behavior \citep{CochranWKH2020a,KarimiD2023a}. The decay parameter ($\alpha$) of the Hawkes model captured the diminishing influence of past events over short time, consistent with patterns expected in induced seismicity driven by transient operational factors. 

To justify the use of the Hawkes model, we performed the fitting of the standard ETAS model to the same catalog. The comparison between the two fits showed a critical distinction: the near-zero magnitude scaling exponent $\alpha=0.0$ of the ETAS model confirmed that triggered sequence rates were largely independent of event magnitudes (Figure~S8). This result supports the hypothesis that external factors, such as mining operations, drive the rate of the temporal evolution of microseismicity compared to the internal stress redistribution. The both models produced very similar fits of the observed seismicity rate as confirmed by the corresponding Akaike information criterion values: -40492.99 for the Hawkes model and -40502.48 for the ETAS model, respectively.

\section{Conclusions} \label{sec:conclusions}

In this study, we presented a statistical approach to analyze induced seismicity in soft rock mining environments, combining spatial and temporal statistical models to provide a deeper understanding of microseismic activity. The integration of the generalized NND method with a mixture model of the two tapered Pareto distributions allowed us to identify and characterize the mircoseismic clusters. By considering the deviations from GR scaling, we analyzed the unique frequency-magnitude characteristics of potash mining seismicity. The temporal Hawkes point process further provided additional insights into the temporal behavior of seismicity, quantifying interplay between the background rate and a self-excitation mechanism. The model demonstrated that induced seismicity is driven primarily by interactions between external operational factors and intrinsic geological responses.

Our findings have both theoretical and practical implications. Theoretically, this study introduced a generalized NND method, a mixture model to describe the frequency-magnitude statistics of events, and applied the Hawkes point process to model mining induced microseismicity. Practically, the results offer further insights for seismic hazard assessments, particularly in mining environments where induced seismicity can pose risks to infrastructure and operational safety.

\begin{datres}
    The data analysis was performed using MATLAB software and the computer scripts used in this study can be downloaded from (\url{https://github.com/rshcherb/GNND_Hawkes}).
    
    The mining microseismicity catalog used in the analysis cannot be publicly shared due to confidentiality restrictions imposed by Nutrien Ltd.
\end{datres}

\section{Declaration of Competing Interests}

The authors acknowledge that there are no conflicts of interest recorded.\footnote{The authors acknowledge that there are no conflicts of interest recorded.}

\begin{ack}
    We would like to thank two anonymous reviewers for their constructive comments and suggestions, which helped to improve the presentation and clarify the results. This research was supported by funding from MITACS and the International Minerals Innovation Institute for the study of mining microseismicity (grant IT25477). R.S. also acknowledges support from the NSERC Discovery grant RGPIN-2020-06424.
\end{ack}


\end{document}